\documentclass[conference]{IEEEtran}
\IEEEoverridecommandlockouts
\usepackage{cite}
\usepackage{amsmath,amssymb,amsfonts,amsthm}
\usepackage{mathtools}
\usepackage{blkarray, bigstrut} 
\usepackage{algorithmic}
\usepackage{graphicx}
\usepackage{textcomp}
\usepackage{xcolor}
\usepackage{bm}
\usepackage{enumitem}
\usepackage{tabu}
\usepackage{makecell}
\usepackage{multirow}
\usepackage{etoolbox}
\makeatletter
\patchcmd{\@makecaption}
  {\scshape}
  {}
  {}
  {}
\makeatother
\theoremstyle{definition}
\newtheorem{theorem}{Theorem}

\newtheorem{definition}{Definition}

\newtheorem*{remark}{Remark}
\def\BibTeX{{\rm B\kern-.05em{\sc i\kern-.025em b}\kern-.08em
    T\kern-.1667em\lower.7ex\hbox{E}\kern-.125emX}}

\begin{document}

\title{Identifying Bridges and Catalysts for Persistent Cooperation Using Network-Based Approach 
}

\author{\IEEEauthorblockN{1\textsuperscript{st} Xingru Chen}
\IEEEauthorblockA{\textit{School of Science} \\
\textit{Beijing University of Posts and Telecommunications}\\
Beijing 100876, China \\
xingrucz@gmail.com}
\and
\IEEEauthorblockN{2\textsuperscript{nd} Feng Fu}
\IEEEauthorblockA{\textit{Department of Mathematics} \\
\textit{Department of Biomedical Data Science} \\
\textit{Dartmouth College}\\
Hanover, NH 03755, USA \\
fufeng@gmail.com}
}
\maketitle

\begin{abstract}
The framework of iterated Prisoner's Dilemma (IPD) is commonly used to study direct reciprocity and cooperation, with a focus on the assessment of the generosity and reciprocal fairness of an IPD strategy in one-on-one settings. In order to understand the persistence and resilience of reciprocal cooperation, here we study long-term population dynamics of IPD strategies using the Moran process where stochastic dynamics of strategy competition can lead to the rise and fall of cooperation.  Although prior work has included a handful of typical IPD strategies in the consideration, it remains largely unclear which type of IPD strategies is pivotal in steering the population away from defection and providing an escape hatch for establishing cooperation. We use a network-based approach to analyze and characterize networks of evolutionary pathways that bridge transient episodes of evolution dominated by depressing defection and ultimately catalyze the evolution of reciprocal cooperation in the long run. We group IPD strategies into three types according to their stationary cooperativity with an unconditional cooperator: the good (fully cooperative), the bad (fully exploitive), and the ugly (in between the former two types). We consider the mutation-selection equilibrium with rare mutations and quantify the impact of the presence versus absence of any given IPD strategy on the resulting population equilibrium. We identify catalysts (certain IPD strategies) as well as bridges (particular evolutionary pathways) that are most crucial for boosting the abundance of good types and suppressing that of bad types or having the highest betweenness centrality. Our work has practical implications and broad applicability to real-world cooperation problems, such as conceiving protocols of steering control and strategy intervention by leveraging catalysts and bridges that are capable of strengthening persistence and resilience.
\end{abstract}


\section{Introduction}

Understanding how cooperation evolves and is sustained is a prominent problem of broad interest and primary significance~\cite{axelrod1981evolution}. Among others, direct reciprocity has been extensively studied using the Iterated Prisoner's Dilemma (IPD) games, where individual behavior is categorized as different strategies~\cite{trivers1971evolution, nowak2006five}. In particular, the so-called Zero-Determinant (ZD) strategies is a set of rather simple memory-one strategies that can unilaterally set a linear relation between their own payoffs and that of their opponent~\cite{press2012iterated, stewart2013extortion}. The finding of such a powerful control over payoffs has greatly spurred new waves of work from diverse fields including network science, computer science, and social science, aiming to shed light on the robustness and resilience of cooperation by means of the natural selection of IPD strategies~\cite{hilbe2013adaptive, chen2022evolutionary, chen2022intricate, chen2023outlearning}.

Prior work on IPD strategies focuses on their ability to foster fairness and cooperation among pairwise interactions. Because of the uncertainty in opponent types, IPD strategies can be optimized in terms of their tolerance, retaliation, and reconciliation of defective moves and also their level of self-recognition and discerning co-players. For example, Grim Trigger (also known as Grudger) retaliates the opponent's defection by turning to defection forever~\cite{banks1990repeated}, and Tit for Tat (TFT) always replicates the opponent's previous move~\cite{axelrod1980effective}. In contrast, Tit for Two Tats (TF2T) can tolerate the co-player's defection not more than twice in a row before taking revenge~\cite{axelrod1980more}. Contrite TFT can also reconcile errors and mistakes in moves with cooperation onwards~\cite{wu1995cope}. Suspicious TFT uses defection as an initial trial of the opponent's type because of the low trust of others~\cite{hilbe2013adaptive}. Collective strategies further take a particular sequence of initial moves to distinguish ``us vs them'' and only cooperate with themselves~\cite{li2009strategy}. Even though this field has been extensively studied with more than 100 common IPD strategies discovered, the framework of IPD games still remains and increasingly becomes an important testbed for combining ideas from artificial intelligence and game theory~\cite{sandholm1996multiagent, bloembergen2015evolutionary, leibo2017multi, harper2017reinforcement, foerster2018learning, lin2022online}. 

A most striking discovery of IPD strategies is that ZD strategies are able to enforce a unilateral linear relationship between their own payoff and their co-player's~\cite{press2012iterated, stewart2013extortion}. By prescribing their particular move conditional on each outcome, ZD players can control the payoff results and even demand an unfair share of the payoffs accrued from interactions. Inspired by this fact, previous studies classify IPD strategies using a dichotomy by their intention of cooperation and ability to reciprocate: partner vs rival strategies~\cite{hilbe2018partners}. These strategies themselves are powerful and form a refined subset of IPD strategies. In this work, we extend prior studies and focus on the morality of strategies based on a simple yet intuitive definition: good strategies that are fully cooperative with an unconditional cooperator (ALLC), bad strategies that are fully defective with an unconditional cooperator, and ugly ones that fall in between. The present classification can lead to a mesoscopic description of cyclic population dynamics in a manner similar to Rock-Paper-Scissors games~\cite{kerr2002local}. 

Although prior work has included a handful of strategies in consideration, it remains largely unclear which types of strategies are pivotal in steering the population away from defection and providing an escape hatch for establishing cooperation. To characterize such transitions between strategies, we consider a network of strategies where evolutionary pathways between them can be evaluated as a direct graph, depending on their ability to disfavor defection and foster the evolution of reciprocal cooperation over the long term.


In our model, individuals play against each other with prescribed IPD strategies and obtain their respective average payoffs from the interactions. Their payoffs, subsequently, will determine their reproductive fitness. To study evolutionary competition, we use a Moran process in a population of finite size and with a significantly low mutation rate~\cite{moran1958random}. The long-term equilibrium of the corresponding evolutionary dynamics can be analytically derived using the approximation method of an embedded Markov chain~\cite{fudenberg2006imitation, antal2009mutation}. Moreover, we can define a directed network where (i) the nodes are the strategies, (ii) the direction of an edge indicates the fixation of the ``target strategy" in a homogeneous population of the ``source strategy", and (iii) the weight of an edge is the ratio of the two fixation probabilities between the pair of strategies. The creation and manipulation of this network allow us to incorporate standard graph measures and algorithms to analyze the functionality of any strategy in the population.

The network-based method helps describe and compare competing IPD strategies. Above all, it can be used to identify essential IPD strategies like Win-Shift, Lose-Stay~\cite{li2011engineering} (the reverse of Win-Stay, Lose-Shift (WSLS)~\cite{nowak1993strategy}) that act as catalysts and transitions between strategies such as from ALLD to Win-Shift, Lose-Stay that act as bridges for recovering cooperation from defection. Their presence plays an important role in the robustness and persistence of cooperation. Our findings share strong similarities with previous studies on the ecological stability and resilience of food webs~\cite{rutledge1976ecological, paine1980food, pimm1991food}.
 
\section{Methods and Model}

\subsection{Payoff structure}

The Prisoner's Dilemma (PD) game is a symmetric game involving two players X and Y, and two actions: to cooperate or to defect. In a one-shot PD game, the four possible outcomes correspond to different payoffs from the focal player's perspective: if both are cooperators, one gets the reward $R$, if a cooperator is against a defector, the sucker's payoff $S$, if a defector is against a cooperator, the temptation $T$, and if both are defectors, the punishment $P$. The game is considered a paradigm for understanding the conflict between self-interest and collective interest as the payoff structure satisfies $T > R > P > S$. 

The iterated Prisoner's Dilemma (IPD) games further assume repeated encounters between the same two individuals and shed insights into the idea of direct reciprocity~\cite{trivers1971evolution}. For any pair of two IPD strategies, denoted by A and B without loss of generality, we can compute the average payoff matrix for their game interactions:
\begin{equation}
\begin{blockarray}{ccc}
 & \text{A} & \text{B}  \\
\begin{block}{c[cc]}
\text{A} & a & b  \bigstrut[t] \\
\text{B} & c & d  \bigstrut[b]\\
\end{block}
\end{blockarray}.
\label{eq:matrix_IPD}
\end{equation}
The pairwise competition dynamics between A and B typically fall into four types~\cite{nowak2006evolutionary}:
\begin{enumerate}[label = (\alph*)]
\item dominance, if 
$a > c$ and $b > d$ or $a < c$ and $b < d$;
\item bistability, if $a > c$ and $b < d$;
\item coexistence, if $a < c$ and $b > d$;
\item neutrality, if $a = c$ and $b = d$.
\end{enumerate}

\subsection{Classification of IPD strategies}

Although hundreds of strategies for the IPD games have been published in the literature, the set is almost negligible compared with the inexhaustible strategy space. To further evaluate these strategies, it was a natural question to ask: how can they be divided into groups?

The strategies can be labeled with their memory lengths which indicate the amount of information they can hold in mind. There are memory-zero strategies whose current actions do not depend on the history of the match. Next comes memory-one strategies remembering only the outcome of the previous round. So on and so forth. A strategy can even have infinite memory, for instance, a Looker-up strategy which always remembers the result of the very first round~\cite{harper2017reinforcement}.

Despite that the classification by memory lengths is plain and simple, it cannot reflect the competitiveness and dominance of strategies in general. Later on, another taxonomy was put forward, further exploring the evolutionary relevance of strategies~\cite{hilbe2018partners}.

\begin{definition}[Nice Strategies]
A strategy is \textbf{nice} if it is never the first to defect $\Leftrightarrow$ if it always cooperates with a cooperator $\Leftrightarrow$ if it always cooperates with itself.
\label{defn:nice}
\end{definition}

\begin{definition}[Cautious Strategies]
A strategy is \textbf{cautious} if it is never the first to cooperate $\Leftrightarrow$ if it always defects against a defector $\Leftrightarrow$ if it always defects against itself.
\label{defn:cautious}
\end{definition}

The equivalence relations are straightforward to show and we omit the proof here.

These two groups of strategies have no intersections. A nice strategy aims for the payoff $R$ while a cautious one refuses to be extorted by others. It is worth noticing that both nice and cautious strategies take up only a measure-zero and hence a negligible portion of the strategy space. There exist other strategies that do not belong to the two groups, for example, the extortionate ZD strategies. 

\begin{remark}
A finer classification based on Definition~\ref{defn:nice} and~\ref{defn:cautious} gives \textbf{partners} and \textbf{rivals} as subsets of nice and cautious strategies, respectively. If a partner's payoff is less than $R$, then the opponent's payoff should also be less than $R$:
\begin{equation}
\pi_X < R \Rightarrow \pi_Y < R.
\end{equation} 
On the other hand, a rival's payoff is always greater than or equal to the opponent's payoff:
\begin{equation}
\pi_X \geq \pi_Y.
\end{equation}  
The ``opposite" of partners and rivals are referred to as \textbf{requiting} strategies and \textbf{submissive} strategies.
\end{remark}

\begin{figure}[htbp]
\centering
\includegraphics[width=0.5\textwidth]{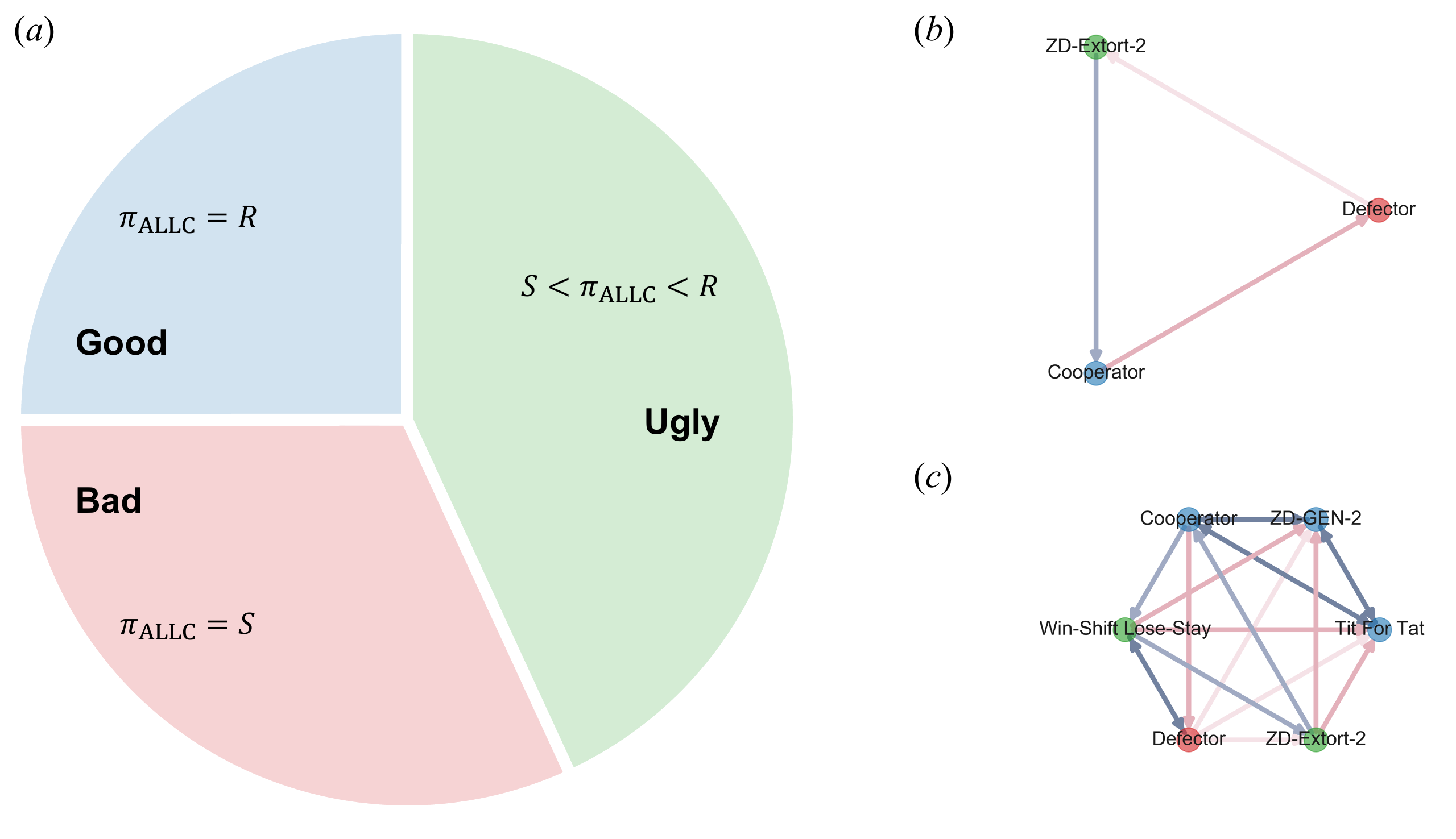}
\caption{The good, the bad, and the ugly. (a) We classify IPD strategies according to their inclination of exploitation against an unconditional cooperator (ALLC): good ones are those who never exploit, bad ones are those who always exploit, and ugly ones are those in between. Unlike prior studies based on pairwise payoff comparison of self vs opponent from the perspective of fairness, our approach offers an intuitive classification that is built on the intrinsic cooperativity of the strategies themselves. (b) and (c) There exists cyclic population dynamics that resembles the classic Rock-Paper-Scissors games in the competition among the good, the bad, and the ugly strategies. A specific cycle of length 3 is given in (b) and a cycle of length 6 is given in (c). The direction of the edges in (b) and (c) indicates greater fixation fixability than the opposite.}
\label{fig1}
\end{figure}

This classification touches on the performance of IPD strategies in the process of evolution. We now introduce our own taxonomy where strategies are divided into three different classes based on the expected payoff of a cooperator as their opponent in the IPD games (see Figure~\ref{fig1}a). The method is simple and works as well as, if not better than, the existing one for identifying the competitiveness of strategies.

\begin{definition}[Good Strategies]
A strategy is \textbf{good} if a cooperator gets an expected payoff $\pi_Y = R$ as its opponent. 
\label{defn:good}
\end{definition}

\begin{definition}[Bad Strategies]
A strategy is \textbf{bad} if a cooperator gets an expected payoff $\pi_Y = S$ as its opponent. 
\label{defn:bad}
\end{definition}

\begin{definition}[Ugly Strategies]
A strategy is \textbf{ugly} if a cooperator gets an expected payoff $S < \pi_Y < R$ as its opponent.
\label{defn:ugly}
\end{definition}

Intuitively, good strategies tend to be friendly to an innocent opponent while bad strategies are determined to exploit the same opponent. It is straightforward to make a comparison between different groups of strategies from the classifications of nice-cautious and good-bad-ugly. For instance, a nice strategy and hence a partner is always a good strategy.  According to~\cite{hilbe2018partners}, successful strategies from the perspective of evolution are either partners or rivals. We can draw a similar conclusion given Definition~\ref{defn:good},~\ref{defn:bad}, and~\ref{defn:ugly}: successful individuals are either good or bad.

\subsection{Evolutionary dynamics}

We use a network-based approach to analyze and characterize networks of evolutionary pathways that bridge transient episodes of evolution dominated by depressing defection and ultimately catalyze the evolution of reciprocal cooperation in the long run. In our model, individuals play against each other using their prescribed IPD strategies (there are $m$ many) and obtain an average payoff from their interactions, $\pi_i$. Their payoffs will further determine their reproductive fitness, $f_i = \exp[\beta \pi_i]$, where $\beta$ is the selection strength. 

To study evolutionary competition, we use a Moran process in a population of finite size $N$: an individual is chosen to reproduce an offspring with probability proportional to their fitness. With probability $\mu$, a mutation occurs and the offspring will randomly choose one of the available $m$ strategies. With probability $1 - \mu$, the offspring is identical to the parent. This newly produced offspring will replace another individual randomly chosen from the entire population.

In the limit of rare mutations where $\mu \to 0$, the fate of a mutant is determined, either reaching fixation or going extinct before the next new mutant arises. Under this assumption, the evolutionary dynamics in finite populations will dwell on a homogeneous population state most of the time, followed by stochastic transitions from one state to another. And the transition between any two population states is determined by the pairwise invasion dynamics (Figure 2). The transition rate can be calculated using the fixation probabilities $\rho_{ij}$, which is the probability that a population of strategy $i$-players is invaded and taken over by a single strategy $j$-player (Figure 2b). Assuming the payoff matrix for $i$ vs $j$ as in~\eqref{eq:matrix_IPD}, the ratio of fixation probabilities is given by
\begin{equation}
\frac{\rho_{ji}}{\rho_{ij}} = \exp\{\beta [\frac{N}{2}(a+b-c-d)-a+d]\}.
\end{equation}

Moreover, the long-term equilibrium of the full evolutionary dynamics for $m$ multiple strategies can be analytically studied using the approximation method of an embedded Markov chain. 

\begin{theorem}[Imitation Processes with Small Mutations]
The $m \times m$ Markov matrix of the $m$ homogeneous states is determined by the transition rates between pairs of strategies. In general, the $ij$th component is
\begin{equation}
m_{ij} = 
\begin{dcases}
\mu\rho_{ij}, & i \neq j \\
1 - \displaystyle\sum_{i \neq j}\mu\rho_{ij}. & i = j
\end{dcases}
\label{eq:Markov}
\end{equation}
In addition, there exists a unique vector $v = [v_1, v_2, \cdots, v_m]$ satisfying 
\begin{equation}
\bm{v}M = M, \qquad \displaystyle\sum_{i = 1}^{m}v_i = 1, \qquad v_i \geq 0, \qquad  \forall i,
\end{equation}
where the $i$th component $v_i$ is the abundance of strategy $i$ after the system becomes stable.
\label{thm:abundance}
\end{theorem}

We use Theorem~\ref{thm:abundance} to compute the stationary distribution of IPD strategies under the limit of a small mutation rate as shown in the Results section.

\begin{figure}[htbp]
\centering
\includegraphics[width=0.5\textwidth]{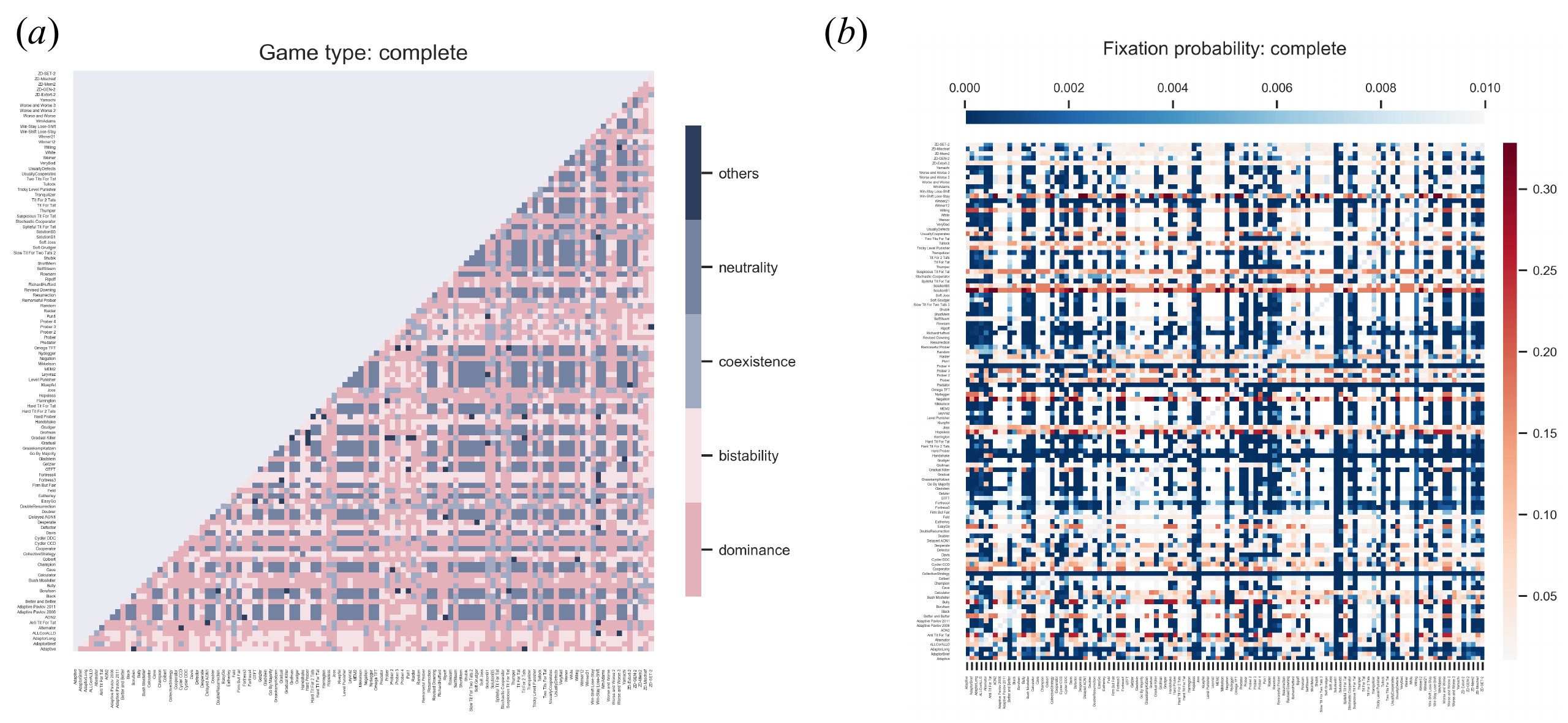}
\caption{Pairwise competition dynamics between common IPD strategies. We consider a representative set of 110 strategies that have been studied in the literature. We first classify the nature of overall game dynamics between any pair of two IPD strategies into five typical scenarios: dominance, bistability, coexistence, neutrality, and others that do not fit into these former four types. We then calculate the fixation probability $\rho_{ij}$ of one single individual with strategy $j$ taking over a resident population playing strategy $i$, the so-called pairwise invasion dynamics. These calculations help us understand the performance of strategies from an evolutionary perspective. Model parameters: population size $N = 100$, and selection strength $\beta = 0.1$.}
\label{fig2}
\end{figure}

We can now consider the $m$ strategies as network nodes and define a directed edge from strategy $i$ to strategy $j$ if the latter has a greater fixation probability and the weight is given by $\rho_{ij}/\rho_{ji}$, and vice versa. In such a way, we obtain a weighted network that consists of different IPD strategies for further identification of catalysts and bridges. We search exhaustively ugly strategies and transitions from or to ugly strategies boosting the abundance of good strategies and suppressing that of bad strategies. There exist cycles in this directed network, a closed directed loop of different IPD strategies that can give rise to cyclic population dynamics of persistent cooperation in a Rock-Paper-Scissors manner.

\section{Results}

Due to the symmetry in which good strategies and AllC each get an average payoff $R$, whether a good strategy can be favored over ALLC in the pairwise competition dynamics is determined by their self-cooperation levels. As such, the presence of good strategies is pivotal to determining the fate of ALLC. On the other hand, good strategies can be viewed as allies of ALLC, bad strategies can be seen as suppressors of ALLC, and ugly strategies can be considered in between the two extremes.  

To characterize and quantify the long-term collective success of good strategies as compared to their counterparts of bad and ugly ones, we study the evolutionary dynamics with a pool of prescribed IPD strategies with rare mutations. In this limit, the population spends most of the time in homogeneous states and the fate of any new mutant, either fixation or extinction, is determined before the next mutant arises in the population. To gain analytical insights, we first identify the type of game interactions between each pair of IPD strategies as shown in Figure 2a. We distinguish four types: dominance (their percentage is $43\%$), bistability ($23\%$), co-existence ($9\%$), neutrality ($24\%$), and a few others that do not fall into these four types. Remarkably, neutrality takes up a substantial proportion. The pairwise fixation probabilities are shown in Figure 2b (noting the dependence on selection strength $\beta$ and population size $N$). Under rare mutations, we are able to analytically calculate the stationary distribution for any given set of IPD strategies under consideration (see Methods and Model section). We confirm a good agreement between analytical results and agent-based simulations. 

We use a directed weighted network to capture all possible evolutionary pathways between any pair of IPD strategies. Each edge can be distinguished by the type of game interactions as shown in Figure 3, and their weight is further given by the ratio of fixation probabilities. The neutrality games, especially between good IPD strategies (Figure 3d), provide an escape hatch for sustaining cooperation and also increasing resilience against perturbations such as invasion attempts by bad strategies.

\begin{figure}[htbp]
\centering
\includegraphics[width=0.5\textwidth]{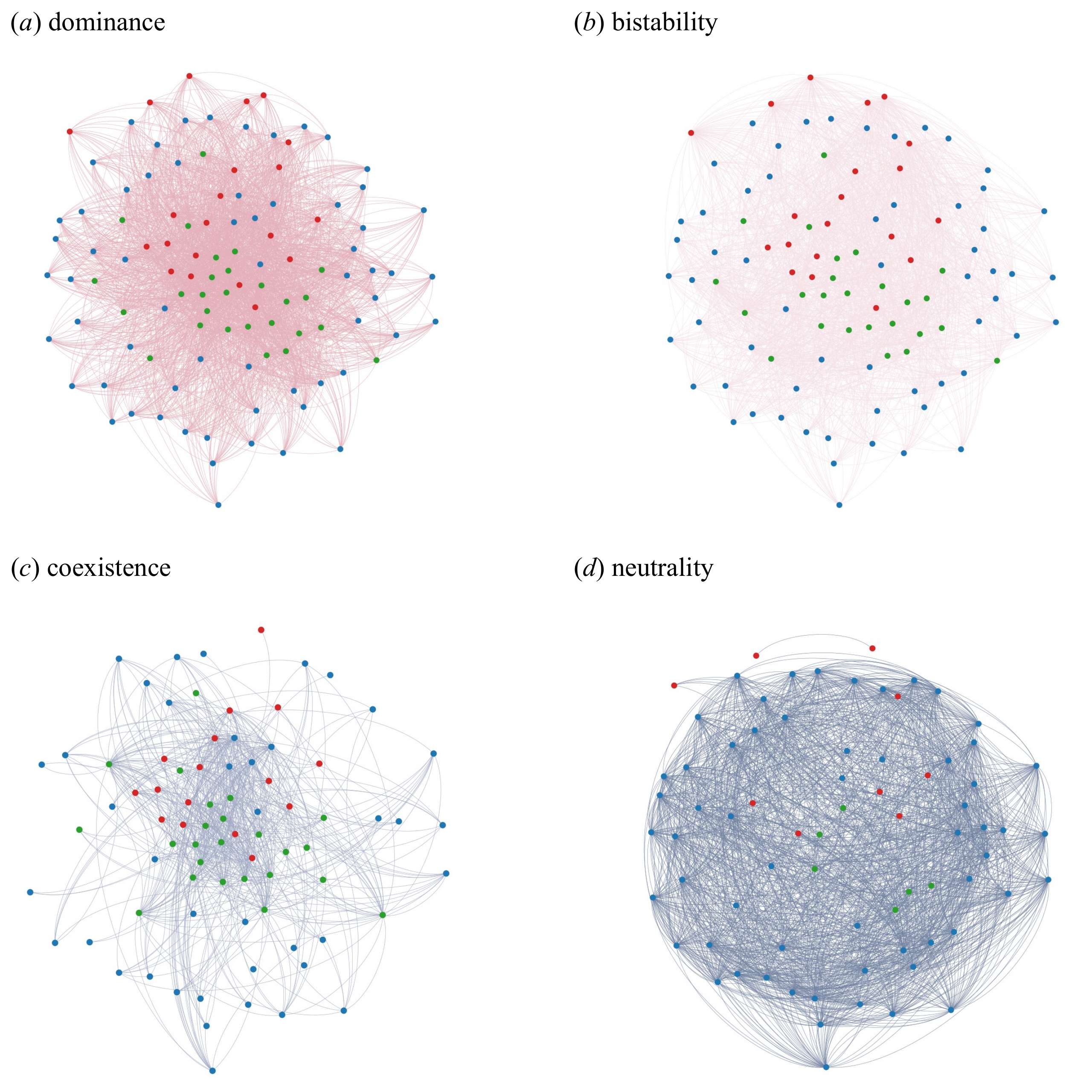}
\caption{Network of IPD strategies arising from pairwise competition dynamics. Each panel shows the subset of edges that are of type (a) dominance, (b) bistability, (c) co-existence, and (d) neutrality, corresponding to the competition types summarized in Figure 2a. This directed and weighted network is constructed based on the pairwise fixation probabilities in Figure 2b. Node color denotes the type of individual node. To visualize with clarity, directionality is not displayed with arrows but using curved edges, which can be read clockwise pointing from a source node to a target node. See main text for details.}
\label{fig3}
\end{figure}

Our network-based approach works for scenarios involving arbitrarily many IPD strategies. In order to get a clear-cut view with an intuitive understanding, we focus on a small sample of 25 IPD strategies, which are either memory-zero (3) or memory-one (22) yet contain the most prominent ones such as TFT, WSLS, and ZD strategies. We use pairwise fixation probabilities (Figure 4a) to compute the stationary abundances of strategies (Figure 4b), and the lump sum abundances of good, bad, and ugly strategies are given in the inset (Figure 4b). We see that good strategies are collectively more successful than the other two types. The network containing all possible pairwise evolutionary pathways is visualized in Figure 4c. The weight of each edge is given by the ratio of fixation probabilities pointing from the one disfavored to the other favored. In the case of two strategies that have equal fixation probabilities, the associated edge is bidirectional with weight one. This network will help us to further identify catalysts and bridges pivotal to the evolutionary success of good strategies. 

\begin{figure}[htbp]
\centering
\includegraphics[width=0.5\textwidth]{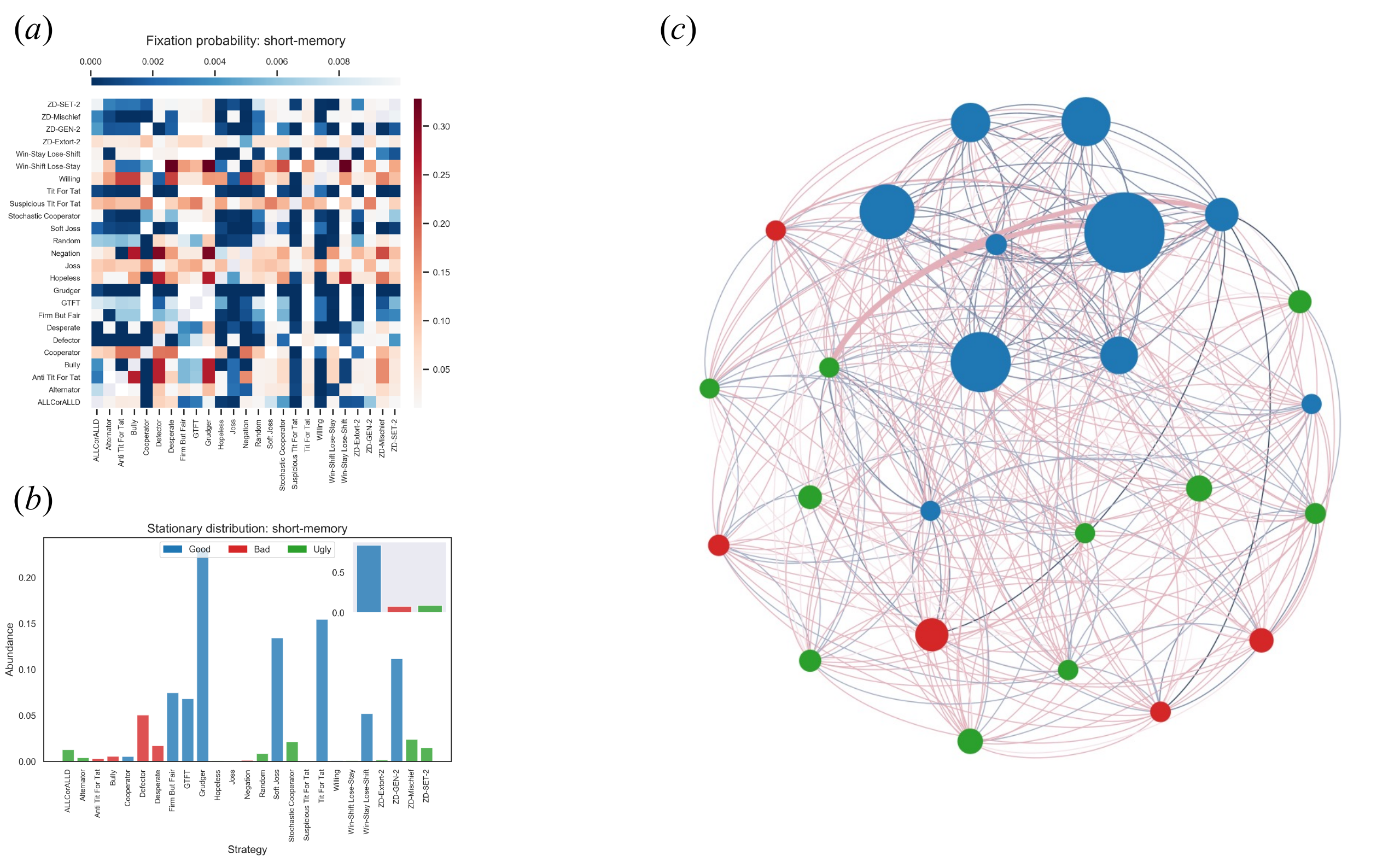}
\caption{Evolutionary dynamics among a small set of 25 IPD strategies for illustration. The heatmap in (a) shows the pairwise fixation probabilities using two different color bars (below $1/N$ vs above $1/N$ as a comparison to neutral evolution). (b) shows the stationary abundance of each sample IPD strategy under consideration. The color of the bar indicates the type of IPD strategies. The network snapshot is shown in (c): node color corresponds to the type, node size is proportional to the abundance of the individual node, and edge color denotes the competition type same as in Figure~\ref{fig3}a. To visualize with clarity, directionality is not displayed with arrows but using curved edges, which can be read clockwise pointing from a source node to a target node. Model parameters: population size $N = 100$, and selection strength $\beta = 0.1$.}
\label{fig4}
\end{figure}

In Figure 5a, we find that Win-Shift, Lose-Stay, an ugly strategy, is the most important catalyst for the evolution of good strategies for varying selection strengths and across different measures. The extortionate ZD strategy (with an extortion factor $\chi = 2$) is the second most important catalyst, followed by the Alternator (an IPD strategy alternating between cooperating and defecting). The most crucial evolutionary pathway, a particular edge in the network, depends also on the specific selection strength $\beta$ and the measure used. For most scenarios (Figure 5b), the directed edge from ALLC to Win-Shift, Lose-Stay is the critical bridge, the presence of which can significantly boost the abundance of good strategies and suppress that of bad strategies while having the highest betweenness centrality among ugly strategies.  

\begin{figure}[htbp]
\centering
\includegraphics[width=0.5\textwidth]{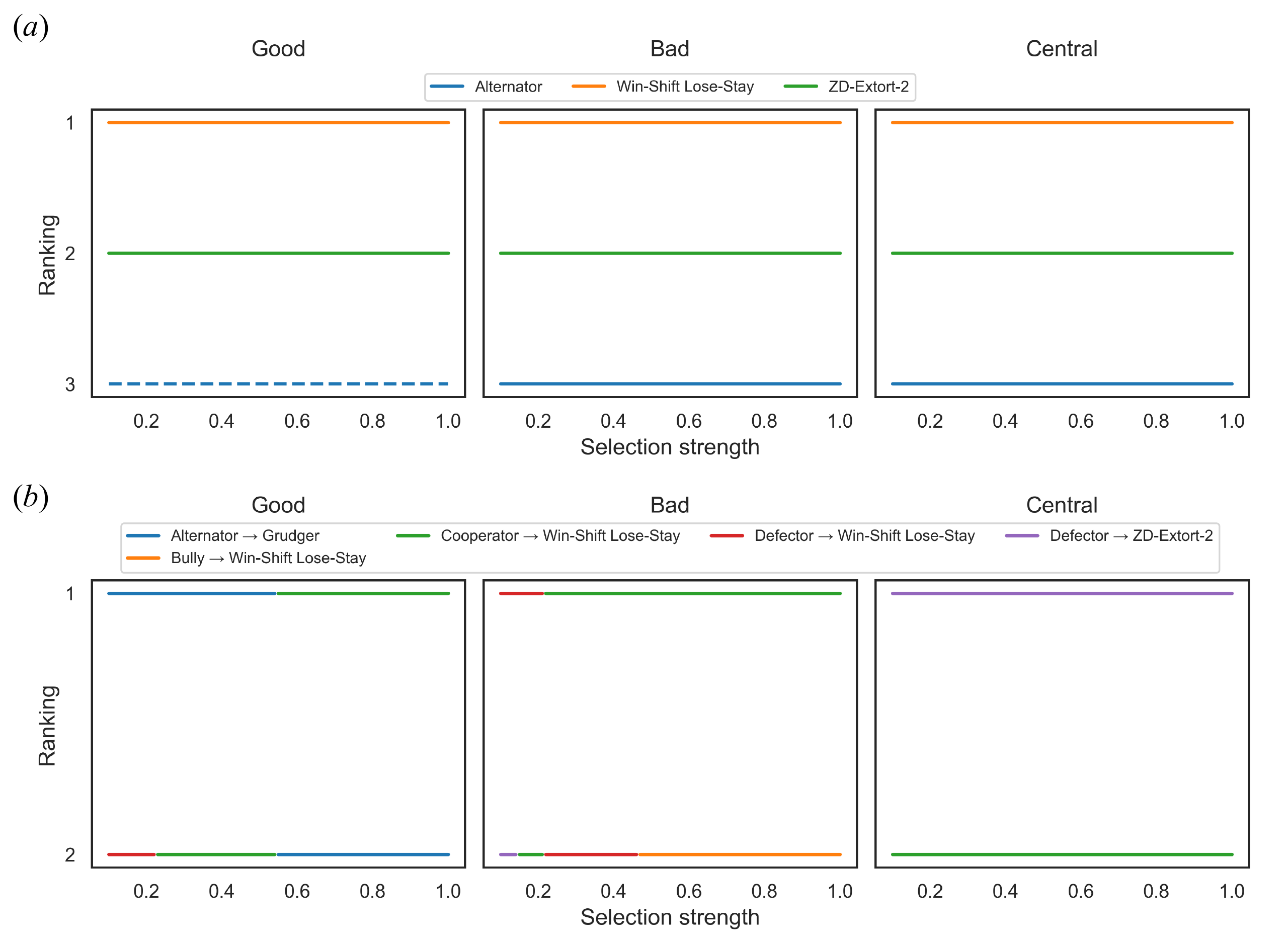}
\caption{Identifying (a) catalysts and (b) bridges that are most crucial in (left panels) boosting the abundance of good strategies, (middle panels) suppressing that of bad strategies, and (right panels) having the highest centrality among ugly strategies and transitions from or to ugly strategies. Results are based on Figure 4.}
\label{fig5}
\end{figure}

\section{Discussion and Conclusion}

Our results demonstrate the very importance of the exact composition of the strategy set when studying natural selection (including but not limited to evolutionary stability and resilience) of particular IPD strategies in population dynamics. Our network-based approach can be used to evaluate and assess the impact of including or excluding an individual IPD strategy on the evolution of good strategies and overall cooperation level. Besides, our method can be applied to steer population dynamics to desired states by adding one or more additional catalysts (and forming bridges that act as evolutionary ``ramps'') that are amenable to external controls. This potential extension is an important insight arising from the present study.

In this work, we consider error-free IPD games where players perfectly implement their intended moves and perfectly observe others' moves. However, noisy games where trembling hands or fuzzy minds are at play are worthy of further investigation~\cite{axelrod2012launching}. It remains an open problem how intelligent players discern intentional deception from innocent errors and mistakes. In particular, the mechanism that they find a common ground notwithstanding different goals and perceptions of fairness is an important and promising area for future work~\cite{dafoe2020open}. 

The present study focuses on fixation probabilities in stochastic dynamics. We emphasize that the evolutionary time scale is another important quantity that should be taken into account. Some evolutionary pathway from one type to another is only possible in theory, that is, the probability of taking over (fixation probability) is non-zero but the expected conditional time for such fixation to occur is exponential under certain circumstances. For example, the fixation time can tend to infinity when the interaction is a snowdrift game as a particular type of co-existence and the selection strength $\beta$ is non-weak~\cite{antal2006fixation}. In light of this, the identification of bridges and catalysts needs to take into account the reasonable requirement of fixation time scales as well. The prior observation of stochastic tunneling in which a third type swiftly takes over a protracted mixture of two competing types could be quite useful in refining the search for the optimal presence of catalysts~\cite{iwasa2004stochastic}.

In the well-studied evolutionary dynamics between ALLC and ALLD where ALLC can never be favored, adding a third type such as TFT or Loners can fundamentally alter the underlying evolutionary dynamics~\cite{hauert2002volunteering}, possibly leading natural selection to favor ALLC over ALLD. Our work generalizes this prior insight to multiple IPD strategies and offers a novel perspective of intervention and control. A targeted suppression and/or promotion of certain subgroup strategies can be achieved by adding or removing certain types of strategies from the pool. Our work highlights that the natural selection of strategies depending on the presence or absence of certain strategies is nontrivial and has broader applications to steering control problems with a broader context~\cite{wang2020steering}. 

In conclusion, we have characterized and compared competing IPD strategies using a network-based approach. Our method can be used to identify essential IPD strategies (e.g. Win-Shift, Lose-Stay) and transitions between strategies (e.g. ALLD to Win-Shift, Lose-Stay) that act as catalysts and bridges for recovering cooperation from defection, and thus their presence plays an important role in the robustness and persistence of cooperation. Our findings resemble some interesting similarities with previous studies on the ecological stability and resilience of food webs~\cite{rutledge1976ecological, paine1980food, pimm1991food}. 

\section*{Acknowledgment}

X.C. gratefully acknowledges the generous faculty startup fund provided by BUPT (No. 505022023). 

\bibliographystyle{unsrt}
\bibliography{ref}

\end{document}